\documentclass[preprint,floats,aps,epsfig,nofootinbib,amssymb]{revtex4}
\usepackage{graphicx,color}
\newcommand{\lsim}{\buildrel < \over {_\sim}}

\begin{document}
\vspace*{-5.8ex}
\hspace*{\fill}{NPAC-12-21}

\vspace*{+3.8ex}

\title{Hidden from View: Neutrino Masses, Dark Matter and TeV-Scale Leptogenesis  in a Neutrinophilic 2HDM}

\author{Wei Chao$^{1,2}$}
\email{chaow@physics.wisc.edu}
\author{Michael J. Ramsey-Musolf${^{1,3}}$ }
\email{mjrm@physics.wisc.edu}
 \affiliation{$^1$ Department of Physics, University of Wisconsin-Madison,Madison, WI 53706, USA\\ 
 $^2$ Shanghai Jiao Tong University, Shanghai 200240, China\\
$^3$California Institute of Technology, Pasadena, CA 91125, USA }

\begin{abstract}
We consider a simple extension of the Standard Model providing dark matter and a TeV-scale
seesaw mechanism that also allows for viable leptogenesis. In addition to the Standard Model
degrees of freedom, the model contains a
neutrinophilic Higgs doublet, a scalar singlet, and six
singlet fermions (including three right-handed Majorana neutrinos) that are charged under a local $U(1)^\prime$ gauge symmetry. 
We show how the  $U(1)^\prime$ charge assignments and the choice of scalar potential can lead to a TeV-scale seesaw
mechanism and $\mathcal{O}(1)$ neutrino Yukawa couplings in a straightforward way.  While this scenario has all the ingredients 
one would expect for significant experimental signatures, including several new TeV scale degrees of freedom,  we find that most distinctive features associated with neutrino mass generation, leptogenesis  and the dark sector are likely to remain inaccessible in the absence of additional lepton flavor symmetries.


\end{abstract}

\draft

\maketitle

\section{Introduction}

Although the Standard Model(SM) is in spectacular agreement with the results of
most terrestrial experiments, it is certainly 
incomplete. Apart from theoretical considerations, such as the
hierarchy problem, origin of electric charge quantization, and the \lq\lq near miss" for gauge unification in the SM, several observations point to the need for physics beyond the Standard Model (BSM).  The observation of neutrino oscillations has revealed
that neutrinos  have non-zero masses and that lepton flavors are
mixed\cite{pdg}. In addition, precisely cosmological observations have
confirmed the existence of non-baryonic cold dark matter:
$\Omega_D^{} h^2 = 0.1123\pm0.0035$ \cite{wmap}. Together with the cosmic baryon asymmetry, 
these  important
discoveries can not be accommodated in the minimal SM without introducing
extra ingredients.

Perhaps the most attractive approach towards understanding the
origin of small neutrino masses is using the dimension-five Weinberg
operator \cite{weinberg}:
\begin{eqnarray}
{1 \over 4 } \kappa_{gf}^{}
\overline{\ell_{Lc}^C}^g\varepsilon_{cd}^{} \phi_d^{} \ell_{Lb}^f
\varepsilon_{ba}^{} \phi_a^{} + {\rm h.c.} \; , \label{weinberg}
\end{eqnarray}
which comes from integrating out some new  super-heavy  particles. A
simple way to obtain the operator in Eq. (1) is through the Type-I
seesaw mechanism \cite{seesawI}, in which three right-handed
neutrinos $N_R$ having large Majorana masses but no SM charges are introduced. Through Yukawa interactions of the $N_R$ with the SM leptons, the
three active neutrinos then acquire tiny Majorana masses as given by  the
Type-I seesaw formula: i.e., the mass matrix of light neutrinos is
given by 
\begin{equation}
M_\nu^{}= - M_D^{} M_R^{-1} M_D^T\ \ \ ,
\end{equation}
where $M_\nu^{}$ is the mass matrix for the light neutrinos; $M_D^{}$ is the
Dirac mass matrix linking the left-handed light neutrinos to
the $N_R$, and $M_R^{}$ is the mass matrix for the $N_R$. Variations of this idea, including
the Type-II \cite{seesawII} and Type-III \cite{seesawIII} seesaw mechanisms have been widely discussed in the literature.

Seesaw mechanisms are among the most natural ways to generate
tiny neutrino masses, and they can be embedded into more fundamental
frameworks such as grand unified theories or string theory. A
salient feature of the seesaw mechanism is that leptogenesis
\cite{Leptogenesis} can work well to account for the
matter-antimatter asymmetry of the Universe. A lepton asymmetry is
dynamically generated by the CP-violating, and out-of-equilibrium
decays of right-handed neutrinos and then converted into a baryon
asymmetry via $(B+L)-$ violating sphaleron
interactions\cite{sphaleron} that exist in the SM. However seesaw
mechanisms typically lose direct testability on the experimental side. A
direct test of seesaw mechanism would involve the detection of these
heavy seesaw particles at a collider or in other neutrino experiments as well as the measurement of their
Yukawa couplings with the electroweak doublets. In the canonical
seesaw mechanism, heavy seesaw particles turn out to be too heavy,
i.e., $\sim 10^{14 }-10^{16}$ ${\rm GeV}$, to be experimentally
accessible.

In view of the prospects for probing new TeV scale physics, it is interesting to ask how one might lower the conventional see-saw scale to an experimentally accessible one. Indeed, there exist many approaches to lower the seesaw scale \cite{Pilaftsisss,early,smirnov,Hannn,tevtype2,Dorsner:2006fx,FileviezPerez:2008sr,Perez:2009mu,FileviezPerez:2012ab}, but these may seem unnatural. In this paper we propose a new TeV-scale seesaw mechanism. A guiding principle in constructing this framework is invoke as few new degrees of freedom as possible within the context of the seesaw mechanism while addressing the cosmic matter content and yielding some degree of testability. In doing so, we seek to explore features that might be generically present in UV complete theories. 

To this end,  
we work in the framework of two Higgs doublet model (2HDM) with a $U(1)^\prime$ gauge symmetry. The second Higgs doublet $H_n$ rather naturally gets a vacuum expectation value (VEV) through its interaction with the SM Higgs boson in the presence of an additional SM singlet scalar $\Phi$. We first show that its VEV can be relatively small. Through an appropriate choice of $U(1)^\prime$ charges for the SM fields as well as the $H_n$, $\Phi$, and $N_R$, we obtain neutrino masses through the standard see-saw mechanism but allowing $\mathcal{O}(1)$ Yukawa couplings between the  light sector and TeV-scale $N_R$. We show that properties of the massive $Z^\prime$ associated with the spontaneously broken $U(1)^\prime$ are compatible with present phenomenological constraints, while the new SM gauge singlets required for anomaly cancellation may provide a viable cold dark matter candidate. Constraints on the model from lepton flavor violation decays are studied.  The model may also generate the matter-antimatter asymmetry of the Universe through flavor-dependent leptogenesis. 

In principle, this scenario contains several ingredients that should lead to experimental signatures:  new TeV-scale degrees of freedom, including the $N_R$, $Z^\prime$, and dark matter fields; $\mathcal{O}(1)$ Yukawa couplings in the neutrino sector, implying the possibility of significant charged lepton flavor violation; and several new scalar degrees of freedom that can mix with the SM Higgs. We find, however, that in the absence of additional lepton flavor symmetries most of the dynamics associated with neutrino mass generation, leptogenesis, and production of the dark matter relic density will be generally difficult to test experimentally. The magnitude of the  neutrino Yukawa couplings directly relevant to the leptogenesis lepton asymmetry must be smaller than necessary for observable signatures in the next generation of charged lepton flavor violation searches, assuming an anarchical Yukawa matrix structure.
The coupling of the $Z^\prime$ boson to SM particles is suppressed by a tiny $Z-Z^\prime$ mixing angle. The mixing of the new scalar singlet with the SM Higgs, which governs the dark matter annihilation cross section, must be sufficiently small to ensure a dark matter relic density consistent with observation, making its impact on Higgs phenomenology at the Large Hadron Collider unlikely to be observable. In the absence of additional flavor symmetries in the lepton sector, the only possible experimental signature of this scenario would be a spin-independent signal in dark matter direct detection searches. Even this signal would not in itself distinguish this scenario from others that lead to a similar signature.  In short, it is possible that new TeV scale physics may be responsible for three of the strongest motivations for extending the SM yet may remain hidden from view.  Moreover, unlike many other TeV scale BSM scenarios that require additional symmetries\footnote{{\em e.g.}, the assumption of minimal lepton flavor violation~\cite{Chivukula:1987py}, Peccei-Quinn~\cite{Peccei:1977hh} symmetry, {\em etc.}} in order to suppress experimental signatures, the present scenario would require adoption new symmetries or structure in order to allow for effects at the observable level.


Our discussion of the model and these features is arranged as follows: In section II we give a brief
introduction to the model. We study neutrino masses, lepton flavor violation and dark matter phenomenology in sections III and  IV. Section V is
devoted to  flavor-dependent leptogenesis. We
summarize in section VI.

\section{The model}
Our purpose is to explain the tiny but non-zero neutrino masses, dark matter and the cosmic baryon asymmetry  {\em via} TeV-scale new physics. To do so in a minimal manner,  we extend the SM with three right-handed Majorana neutrinos $N_R^{}$, one extra Higgs doublet $H_n^{}$ and one scalar singlet $\Phi$. From the standpoint of the see-saw mechanism,  sufficiently small Dirac masses $M_D$ for TeV scale $M_R$ and $\mathcal{O}(1)$ Yukawa couplings can arise if the VEV of $H_n^{}$ is small and solely responsible for a non-vanishing $M_D$. 
Ensuring that the SM Higgs doublet $H$ does not contribute to $M_D$ in this case requires imposition of an additional symmetry. Such a symmetry can be a  global $U(1)^\prime$ symmetry that is broken both spontaneously and explicitly (softly)\cite{musolf,ernestma, dirneu, zere}, a discrete flavor symmetry \cite{z2haba, girmus,Haba:2011fn,Haba:2011yc,Cogollo:2010jw} or a local $U(1)^\prime$ symmetry. In Table~\ref{aaa} we classify variations of the possible $U(1)^\prime$ extensions, including the relevant particle content and  $U(1)^\prime$ charges.  We distinguish two possibilities for the global $U(1)^\prime$, corresponding to whether or not the additional doublet is charged under this group. In the case of the local $U(1)^\prime$ we denote the additional SM gauge singlet fermions required for anomaly cancellation as $\Psi$. 

As we discuss below, the local $U(1)^\prime$ and type-II global $U(1)^\prime$ also admit a dark matter candidate.  In the latter instance, the pseudo-Goldstone boson associated with spontaneous symmetry breaking is the dark matter particle, while explicit, soft breaking is needed to provide a dark matter mass. A simpler version that includes only the SM doublet $H$ and singlet $\Phi$ and that does not consider neutrino masses has been  studied in Ref.~\cite{musolf}. In the present context,  the generation of the neutrino Dirac mass in the type-II global $U(1)^\prime$ entails an additional, explicit breaking of the symmetry as one may see from the set of possible Yukawa interactions:
\begin{eqnarray}
-{\cal L}_{\rm Y}^{} = \overline{\ell_L^{}} Y_{e ij}^{} H^{} E_R^{}+ \overline{\ell_L^{i}} { Y}^{}_{\nu ij} \tilde H N_R^{j}
+ \overline{\ell_L^{i}} \lambda^{}_{ij} \tilde H_n^{} N_R^{j} + {\rm
h.c.} \; . \label{lyukae}VEV
\end{eqnarray}
Given the $U(1)^\prime$ charges of the $N_R$ and $H$, the second term explicitly breaks the symmetry as does the third term if $H_n$ is uncharged. For the type-II global $U(1)^\prime$, for which one must include an explicit symmetry-breaking term in the potential to generate the dark matter mass, it would be unreasonable not to include both of the symmetry breaking terms in Eq.~(\ref{lyukae}) as well. In this case, one would have to explain the disparity  between a tiny coupling ${Y}_\nu$ associated with the SM Higgs vacuum expectation value (VEV) and an $\mathcal{O}(1)$ $\lambda$ in the presence of the tiny  VEV of the neutral component of $H_n$. Consequently, we do not consider this scenario further. 

In contrast, the type-I global $U(1)^\prime$ allows one to forbid the second term in Eq.~(\ref{lyukae}), while the third term needed for the TeV-scale seesaw mechanism is allowed. However, the most general scalar potential also includes operators that are invariant under the new $U(1)^\prime$ but also provide a mechanism for decay of the pseudo-Goldstone boson, thereby precluding its viability as a dark matter particle. Consequently, we focus our attention on the local $U(1)^\prime$ scenario, for which both a viable dark matter particle and TeV-scale seesaw mechanism appears possible without the introduction of explicit symmetry-breaking terms. 

For the local $U(1)^\prime$ case, the neutral component $H_n^0$ gets a small vacuum expectation value(VEV) through its coupling with the SM Higgs and is solely responsible for the origin of Dirac neutrino mass matrix. Three new SM gauge singlet $\psi_L^{}$ are introduced to cancel the anomalies. The most general scalar potential is then  given by
\begin{table}[htbp]
\centering
\begin{tabular}{c|c|c|c }
\hline \hline ~~~~~~~~~~ &  Local  $ U(1)^\prime $ & Type-I global $U(1)^\prime$  
&Type-II global $U(1)^\prime$ \\
\hline$H $ & $0$ & $0$ & $0$  \\
\hline$H_n^{} $ & $1$ & $1$ & $0$  \\
\hline$\Phi $ & $1$ & $1$ & $1$  \\
\hline$N_R$ & $1$ & $1$ & $1$  \\
\hline$\Psi $ & $1$ &  $\times$ & $\times$  \\
\hline$Z_2^{}$ & $\surd$& $\times$  & $\times$  \\
\hline SB & $\times$ & $\surd$  &  $\surd$ \\
\hline DM & $\Psi$ & $\times $  &  $\delta$ \\
\hline \hline
\end{tabular}
\caption{ Quantum numbers of  fields under local $U(1)^\prime$, type-I and type-II global $U(1)^\prime$ symmetry. Here 
SB denotes explicit symmetry breaking term, DM denotes possible dark matter candidate in this scenario.  }\label{aaa}
\end{table}
\begin{eqnarray}
V&=& - m_1^{2} H^\dagger H^{} + m_2^{2} H_n^\dagger H_n^{} - m_0^2
\Phi^\dagger \Phi + \lambda_0^{} (\Phi^\dagger \Phi)^2 +
\lambda_1^{} (H^\dagger H)^2 + \lambda_2^{}
(H_n^\dagger H_n^{})^2   \nonumber   \\
&&+ \lambda_3^{} (H^\dagger H)(H_n^\dagger H_n^{}) + \lambda_4^{}
(H^\dagger H_n^{} ) (H_n^\dagger H ) + \lambda_5^{} (\Phi^\dagger
\Phi ) (H^\dagger H )\nonumber \\
\label{eq:vscalar}
&& + \lambda_6^{} (\Phi^\dagger \Phi ) ( H_n^\dagger H_n^{} )+
\left[ \lambda_{n}^{}  \Phi ( H_n^{\dagger  } H) + { \rm h.c.
} \right] \; .
\end{eqnarray}
Here we assume $\lambda_n^{}$ is a real parameter with positive mass dimension\footnote{Note that this operator is multiplicatively renormalized logarithmically and is, therefore, technically natural.}. In the case of the type-II global $U(1)^\prime$ scenario, the corresponding operator breaks a $Z_2$ symmetry that would otherwise guarantee stability of the possible dark matter candidate. 
It should also be noted that the interaction term, $(H^\dagger H_n^{} ) (H^\dagger H_n^{}) + {\rm h.c.}$, is forbidden by the $U(1)^\prime$ symmetry.
In what follows, we define  $H=(h^+, ~(h+i A + v_1^{} ) /\sqrt{2})^T$,  $H_n^{}=(\rho^+, ~(\eta+i \delta + v_2^{} ) /\sqrt{2})^T$ and $\Phi= (\phi + i \varphi^{} + v_0^{}) /\sqrt{2}$.  

After imposing the conditions for the VEVS $v_j$ to yield an extremum, one
has
\begin{eqnarray}
\label{eq:cpevenVEVs}
v_0^{2} \approx { 4 m_0^2 \lambda_1^{} -2 m_1^2 \lambda_5^{} \over 4
\lambda_0^{} \lambda_1^{} - \lambda_5^{2} }\; ,
\hspace{1cm } v_1^2 \approx { 4 m_1^2 \lambda_0^{} -2 m_0^2
\lambda_5^{} \over 4 \lambda_1^{} \lambda_0^{} -
\lambda_5^{2} } \; , \hspace{1cm} v_2^{} \approx-{\lambda_n^{} v_0^{}
v_1^{} \over \sqrt{2} m_2^2}\; .
\end{eqnarray}
It is clear that $v_2^{}$ is suppressed by $m_2^2$. By setting $m_2^{2} $ to be sufficiently large or $\lambda_n$ to be tiny, $v_2^{}$ can be of order
several MeV -- the salient feature of our model. In the basis $(h, ~\eta, ~\phi)$  we then derive the mass-squared matrix for the CP-even scalars
scalars
\begin{eqnarray}
M_{\rm CP ~even}^2= \left(\begin{array}{ccc} m_{hh}^2  & 
 v_1^{} v_2^{} (\lambda_3^{} + \lambda_4^{} ) +{
 \lambda_n^{}  v_0^{} \over  \sqrt{2}}&  v_1^{} v_0^{}
\lambda_5^{} +{ \lambda_n^{} v_0^{} \over   \sqrt{ 2 } } \\
\clubsuit & m^2_{ \eta \eta } &v_2^{} v_0^{}
\lambda_6^{} +{ \lambda_n^{} v_1^{} \over   \sqrt{ 2 } } \\
\clubsuit & \clubsuit & m_{ \phi \phi }^2
\end{array} \right) \; ,
\end{eqnarray}
where
\begin{eqnarray}
m_{hh}^2 = 2v_1^2 \lambda_1^{} -{\lambda_n^{} v_0^{}
v_2^{} \over
\sqrt{2} v_1^{}} \; , \hspace{1cm}
m_{ \eta \eta }^2 =  2v_2^2 \lambda_2^{}-{\lambda_n^{}
v_0^{} v_1^{} \over 
\sqrt{2} v_2^{}} \; ,\hspace{1cm}
m_{ \phi \phi }^2  =  2v_0^2 \lambda_0^{}-{\lambda_n^{}
v_1^{} v_2^{} \over \sqrt{2} v_0^{}} \; ,
\end{eqnarray}
and where \lq\lq $\clubsuit$" symbol indicates the entry $(M^2)_{jk}$ is equal to $(M^2)_{kj}$.
We also derive the mass matrix for the $CP-{\rm odd}$ scalars in the basis
$(A, ~ \delta , ~ \varphi^{})$:
\begin{eqnarray}
M_{\rm CP ~ odd}^{2} = \left( \begin{array}{ccc}  m_{AA}^2 & -{ v_0^{}
\lambda_n^{} \over  \sqrt{2}} &- { v_2^{} \lambda_n^{} \over 
\sqrt{2} } \\ \clubsuit & m_{\delta\delta}^2 &+ { v_1^{} \lambda_n^{}
\over \sqrt{2 }} \\ \clubsuit & \clubsuit & m_{\varphi \varphi}^2
\end{array} \right) \; ,
\end{eqnarray}
where
\begin{eqnarray}
m_{AA}^2 &=& {\lambda_n^{} v_0^{} v_2^{} \over 
\sqrt{2} v_1^{}}\; , \hspace{1cm}
m_{\delta\delta }^2 = {\lambda_n^{} v_0^{} v_1^{} \over 
\sqrt{2} v_2^{}}\; , \hspace{1cm}
m_{\varphi \varphi}^2 ={\lambda_n^{} v_1^{} v_2^{} \over 
\sqrt{2} v_0^{}}\; .
\end{eqnarray}
It is straightforward to see that  that there is only one  massive CP-odd scalar with squared mass eigenvalue: $ \lambda_n^{} (v_0^2 v_1^2 + v_1^2 v_2^2+v_0^2 v_2^2) (v_0^{} v_1^{} v_2^{} )^{-1}/\sqrt{2}$. The other two are would-be Goldstone bosons eaten by the  gauge fields $Z$ and $Z^\prime$, respectively. 

Before proceeding, we make additional comparisons with the global $U(1)^\prime$ scenarios. As indicated earlier, the spontaneous breaking of a global $U(1)^\prime$ symmetry would lead to a massless Goldstone boson, a possibility that is severely constrained by big bang nucleosynthesis~\cite{bigbang} and  observations of Bullet Cluster galaxies~\cite{bullet}. Hence, such the global $U(1)^\prime$ symmetry must be explicitly broken, which may be accomplished by adding following terms to the Higgs potential~\cite{musolf}
\begin{eqnarray}
V_\Delta^{}= { \Delta_1^{} }  \Phi^2 + \Delta_2^{}  \Phi + {\rm h.c.} \; . \label{soft}
\end{eqnarray} 
Note that the operators in $V_\Delta^{}$ close under renormalization. Though $\Delta_1^{}$ and $\Delta_2^{}$ are in general complex parameters, we set them to be real for simplicity.  Adding Eq. (\ref{soft}) to  Eq.~(\ref{eq:vscalar}) and imposing the minimization conditions ,  one has
\begin{eqnarray}
M_{\rm CP~ even}^{\prime 2} = M_{\rm CP~ even}^{ 2} + \Delta M^2\; , \hspace{0.5 cm}  M_{\rm CP~ odd}^{\prime 2} = M_{\rm CP~ odd}^{ 2} - \Delta M^2 \; ,
\end{eqnarray}
where 
\begin{eqnarray}
\Delta M^2 =\left(  \begin{array}{ccc} 0 & 0 & 0 \\ 0 &0  &0 \\ 0 & 0 & {2 \Delta_1}\end{array}\right) \; .
\end{eqnarray}
The VEVs of Higgs fields in this case are quite similar to that in Eq.~(\ref{eq:cpevenVEVs}), only up to the replacement $m_0^2 \rightarrow m_0^2 + \Delta_1^{}$. 

Compared with the local $U(1)^\prime$ case,  one does not require the extra fermion singlets to cancel anomalies. However,   one does need to add explicit $U(1)^\prime$ symmetry breaking terms to the Higgs potential by hand. As discussed above, there then exists no reason not to include the explicit $U(1)^\prime$-breaking terms in the $\mathcal{L}_Y$ as well.  For the local $U(1)^\prime$ scenario, the extra fermion singlet can be dark matter candidate, as we discuss in section IV. For the type-I global $U(1)^\prime$ scenario, there is no dark matter candidate. For the type-II global $U(1)^\prime$ scenario, the imaginary part of the scalar singlet can be dark matter candidate \cite{musolf}, but  the $U(1)^\prime$ symmetry in this case is only responsible for the stability of $\phi_a$, and it has nothing to do with the VEV of the new Higgs doublet.

As with any local $U(1)^\prime$ scenario, one must consider $Z$$-$$Z^\prime$ mixing and constraints from electroweak precision observables.  In the present instance, the kinetic term $(D_\mu H_n)^\dagger( D^\mu_{} H_n^{})$ contributes to the $Z$$-$$Z^\prime$ mixing  at both the tree level and one-loop level.  Effects of the latter are characterized by the mixing tensor $\Pi^T_{ZZ^\prime}$ given in the Appendix. Taking both contributions into account, the mass eigenvalues of $Z $ and $Z^\prime$ are then
\begin{eqnarray}
M_Z^{2}& =& {1 \over 4 }c^2 (g^2 + g^{\prime 2}) (v_1^2 + v_2^2) + s^2 g^{\prime \prime 2} (v_0^2 + v_2^2 ) - cs \left(g^{\prime \prime} \sqrt{g^2 + g^{\prime 2}} v_2^2 + 2 \hat \Pi_{ZZ^\prime}^T \right)\; , \\
M_{Z^\prime}^2 &=&     {1 \over 4 }s^2 (g^2 + g^{\prime 2}) (v_1^2 + v_2^2) + c^2 g^{\prime \prime 2} (v_0^2 + v_2^2 ) + cs \left(  g^{\prime \prime} \sqrt{g^2 + g^{\prime 2}} v_2^2 + 2 \hat \Pi_{ZZ^\prime}^T\right)\; ,
\end{eqnarray}
where $c=\cos \theta $ and $s=\sin \theta$, with
\begin{eqnarray}
\tan 2\theta = { 4 v_2^2 g^{\prime\prime} \sqrt{g^2 + g^{\prime 2}} +8\hat \Pi_{ZZ^\prime}^{T} \over 4 g^{\prime \prime 2} (v_0^2 + v_2^2 ) - (g^2 + g^{\prime 2}) (v_1^2 + v_2^2 )} \; .
\end{eqnarray} 
The mixing matrix that appears in the change of the basis form $(A_{3\mu}^{}, ~B_\mu ,~B^\prime_\mu)$ to  $(Z_\mu,~ A_\mu, ~ Z_\mu^\prime )$ is given by
\begin{eqnarray}
\left( \matrix{Z_\mu \cr A_\mu\cr Z_\mu^\prime}\right)=\left(\matrix{c & 0 & -s \cr 0 & 1 & 0 \cr  s & 0 & c} \right) \left(\matrix{c_w^{} & -s_w^{} & 0 \cr s_w^{} & c_w^{} & 0 \cr 0 & 0 &1 } \right) \left(  \matrix{A_{3\mu}^{} \cr B_{\mu} \cr B^\prime_{\mu}}\right)\; ,
\end{eqnarray}
where $(c_w, ~s_w)\equiv (\cos \theta_W, ~\sin\theta_W) $ with $\theta_W^{}$ the standard model weak mixing angle.  Phenomenological constraints typically require the $Z$-$Z^\prime$ mixing angle $\theta$ to be less than $\sim 1 - 2 \times 10^{-3}$ \cite{thetax} and the mass of extra neutral gauge boson to be heavier than $865\sim 910 ~{\rm GeV}$\cite{zpmass}. A suitable mass hierarchy and mixing between $Z$ and $Z^\prime$ are maintained by setting  $v_0^{}$ relatively large( i.e., $v_0^{}>1\, {\rm TeV}$ ),   $v_2^{} \sim 1 ~{\rm MeV}$ and $m_{\eta, \delta} \sim 150~{\rm GeV}$.  However, it should be mentioned that, $Z^\prime$ in our model only couples to the BSM fields such that constraint given above could be relaxed. The masses of $W^\pm$ are $ 1/2 g\sqrt{v^2 + v_2^2}$. It is straightforward to determine that for this choice of parameters, the constraint on the model from the $\rho $ parameter
$
\rho \equiv {M_W^{} \over M_Z^{} \cos \theta_w^{} }\approx 1 \pm 0.001
$
 is maintained as the constraint on $Z-Z^\prime$ mixing is fulfilled.  


\section{Neutrino masses and lepton flavor violations}
Tiny Majorana neutrino masses arise naturally in this framework, as one may observe from Eqs.~(\ref{lyukae},\ref{eq:cpevenVEVs}). We now set $Y_\nu=0$ as required by the $U(1)^\prime$ symmetry and charge assignments. 
The entries in Dirac neutrino mass matrix are proportional to $\lambda_{ij}\, v_2^{}$, such that they may be at the  ${\rm MeV}$ scale while  keeping  $\lambda\sim {\cal O} (1) $.  In this case we only need TeV scale right-handed Majorana neutrinos to suppress the active neutrino masses to the eV  scale. 

The mass term for $N_R$ requires introduction of the dimension-five operator
\begin{equation}
\frac{1}{\Lambda} \Phi^2  \overline{N_R^{}} N_R^C + {\rm h.c}\ \ \ ,
\label{eq:nrmass}
\end{equation}
where the scale $\Lambda$ is presumably associated with integrating out additional heavy fields. For example, one may
make the theory renormalizable by introducing three extra fields $S_L^{}$ that are
gauge singlets and uncharged under the $U_{}(1)^\prime$ symmetry. Their mass term as well as Yukawa interactions with the right-handed neutrinos can be
written as
\begin{eqnarray}
\overline{S_L^{}}Y_N^{}  \Phi^\dagger N_R^{}  + \overline{S_L^C}
M_S^{} S_L^{} + {\rm h.c.} \; .
\end{eqnarray}
Integrating out heavy fields $S_L^{}$, gives $1/\Lambda\sim Y_N^T Y_N/M_S$, yielding the 
right-handed neutrino mass matrix
written as
\begin{eqnarray}
M_R^{} =- v_0^2 Y_N^{} M_S^{-1} Y_N^T \; . \label{rneu}
\end{eqnarray}
Tiny, non-vanishing Majorana masses of active neutrinos can be obtained by integrating out heavy Majorana neutrinos:
\begin{eqnarray}
M_\nu^{} = v_2^2  \lambda M_R^{-1} \lambda^T = v_2^2 v_0^{-2} \lambda (Y_N^T)^{-1} M_S^{} Y_N^{-1} \lambda^T \; .
\end{eqnarray}
Setting ${\cal O}(v_2^{})\sim 1 ~\mathrm{ MeV} $, ${\cal O} (v_0^{}) \sim 1~ \mathrm{TeV}$ and $ {\cal O} ~(M_S^{}) \sim
100~\mathrm{GeV} $, then electronvolt scale  neutrino masses only require $
\lambda$ and $Y_N$ to be $\mathcal{O}(1)$.

The structure of the Yukawa matrix $\lambda$ is of course constrained by the results of neutrino oscillation studies. Below we provide an illustrative example that is consistent with these constraints. In addition, the Yukawa interactions in Eq.~(\ref{lyukae}) lead to non-conservation of charged lepton flavor at the one-loop level. Given the TeV scale masses for the $N_R$ and $\mathcal{O}(1)$ Yukawa couplings $\lambda$ one might anticipate observable charged lepton flavor violation (CLFV). At present,  the most stringent constraint arises from the non-observation of the decay $\mu \rightarrow e + \gamma$. The associated branching ratio is given by
\begin{eqnarray}
BR(\mu \rightarrow e + \gamma) = {3 e^2 \over 64 \pi^2 G_F^{2}}|{\cal A }|^2 \left(1- {m_e^2 \over m_\mu^2 } \right)^3 \; ,
\end{eqnarray}
where the amplitude $\mathcal{A}$ arises at one-loop order and depends on the mass of the $\eta$ scalar and  masses $M_i$ of the TeV-scale $N_R$ as \begin{eqnarray}
{\cal A } = {\lambda_{ e i}^{} \lambda^*_{ \mu i} \over 12(M_i^2 - m_\eta^2)} \left\{2 + {9 m_\eta^2 \over M_i^2 -m_\eta^2 } + 6\left({m_\eta^2 \over M_i^2 -m_\eta^2 }\right)^2 - {6 M_i^4 m_\eta^2 \over (M_i^2 - m_\eta^2 )^3} \ln\left (  {M_i^2 \over m_\eta^2}\right)\right\}\; .
\end{eqnarray}
Here, a sum over $i$ is assumed.

Looking to the future, experiments searching for  the $\mu-e$ conversion in nuclei with competitive sensitivity are planned at Fermilab and J-PARC. For this process, the dominant contribution arises from the exchange of a virtual photon that couples to a loop-induced $\mu-e$ charge radius operator. Nuclear coherence enhances this contribution by $\sim Z^2$ over that associated with the magnetic dipole operator, while 
 the $Z^0$-exchange dipole and charge-radius conversion amplitude is suppressed by a factor of $m_\mu^2 / m_Z^2$.  Retaining only the leading contribution yields the branching ratio \cite{cai, lfvnumber}
\begin{eqnarray}
BR_{\mu \rightarrow e }^A = R^0_{\mu \rightarrow e } (A) \left|1 + {\tilde{g}_{LV}^p V^{p} (A ) \over A_R^{} D(A)}  + {\tilde g _{LV}^n V^n (A) \over A_R^{} D(A)}\right|^2 BR(\mu\rightarrow e \gamma) \; ,
\end{eqnarray}
with
\begin{eqnarray}
R_{\mu\to e}^0 (A)& =& {G_F^2 m_\mu^5 \over 192 \pi^2 \Gamma_{capt}^A} |D(A)|^2 \; ,
\end{eqnarray}
and
\begin{eqnarray}
\tilde g^p_{LV} &=&2 g_{LV}^{u} + g_{LV}^d \; , \hspace{1cm } \tilde g_{LV}^n =  g_{LV}^{u} +2  g_{LV}^d  \; , \hspace{1cm}
A_R^{} ={\sqrt{2} \over 8} {{\cal A } \over G_F^{} m_\mu^{}} \;  ,  \nonumber \\
g_{LV}^q&= &-{s_w^2 \over 72 \pi^2 } {m_W^2 \over M_i^2 -m_\eta^2  } Q_q^{} \lambda_{ei}^{} \lambda_{\mu i}^*\left[2 + { 3  M_i^2  \over  M_i^2 - m_\eta^2 } + 6\left( M_i^2 \over M_i^2 -m_\eta^2\right)^2  - 6 \left( M_i^2 \over M_i^2 -m_\eta^2\right)^3 \ln {M_i^2 \over m_\eta^2 }\right] \; , \nonumber 
\end{eqnarray}
where $D(A)$, $V^p(A)$ and $V^n (A)$ are overlap integrals as a function of atomic number \cite{lfvnumber}.
\begin{figure}[h]
\begin{center}
\includegraphics[width=11cm,height=8 cm,angle=0]{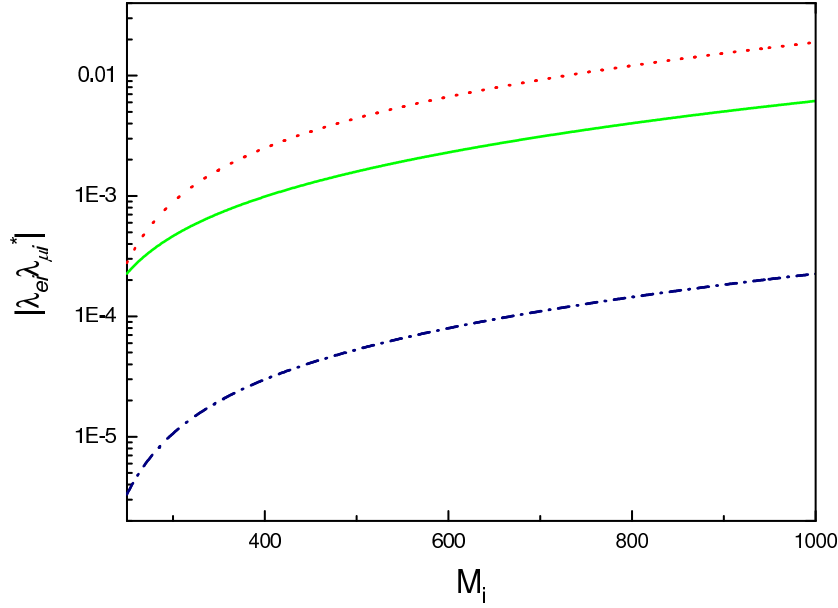}
\end{center}

\caption{ $|\lambda_{\mu i}^* \lambda_{e i}^{}|$ as function of $M_i$ ( assumed here to be degenerate ) constrained by charged lepton flavorviolating processes. The solid line is the constraint from the null result for $\mu\to e \gamma$~\cite{Adam:2011ch}. The dotted line is the constraint of the current null result for $\mu-e $ conversion in $^{197}_{79} {\rm Au}$. The dot-dashed line gives the reach of future $\mu-e$ conversion searches, assuming a sensitivity to the conversion branching ratio of $10^{-18}$\cite{mutoe} .  } \label{fig0}
\end{figure}

The current experimental upper bounds for the $BR(\mu\rightarrow e \gamma )$ and $\mu-e$ conversion  in $^{197}_{79} {\rm Au}$ are $1.2\times 10^{-11}$ and $ 7.0 \times 10^{-13}$ \cite{lfvnumber,Adam:2011ch,mutoe},  respectively. To illustrate the impact of these present and prospective limits, we assume that the heavy neutrinos are degenerate and  plot in Fig. \ref{fig0}, constraints on $|\lambda_{\mu i}^* \lambda_{e i}^{}|$ as a function of $M_i$
implied by $\mu\rightarrow e \gamma$ ( the solid line ) and $\mu-e$ conversion (the dotted line) processes. The dot-dashed line is the possible constraint from the future $\mu-e$ conversion experiments \cite{mutoe}.  By assuming $M_i^{} \sim 300 ~{\rm GeV}$, $m_\rho^{} \sim 150 ~ {\rm GeV}$, we obtain an upper bound for the $|\lambda_{ e  i} \lambda^*_{ \mu i}|$ of roughly $4\times 10^{-4}$.  As a result, $\lambda^{}\sim{\cal O}( 10^{-2})$. It can be found from the figure that the current bounds on $\mu-e$ conversion do not provide significant constrains on the parameter space. On the other hand, the sensitivities of prospective future $\mu-e$ conversion searches \cite{mutoe} exceed that of the MEG experiment.

\section{Dark Matter}

The fact that  about $23\% $ of the Universe is made of dark matter has been firmly established, while the nature of  the dark matter still eludes us. A weakly interacting massive particle (WIMP) is a promising dark matter candidate, since the WIMP relic density can be naturally near the experimental observed value for a WIMP mass  around the electroweak scale.  In our model  the fields $\psi_L^{}$, introduced to cancel anomalies of the $U(1)^\prime$ gauge symmetry (one for each $N_R$), provide a WIMP candidate. The $\psi_L^{}$ mass can be generated through a dimension-5 effective operator, just in the similar way as that for the right-handed neutrinos.
\begin{eqnarray}
{ 1 \over \Lambda}  \tilde{Y}_{ij}\Phi^2 \overline{\psi_{Li}^C} \psi_{Lj}^{} \; ,
\end{eqnarray}
where $\Lambda$ is a cutoff scale.  After the spontaneously broken of the $U(1)^\prime$, $\psi$ gets non-zero mass
\begin{eqnarray}
\left( M_\psi \right)_{ij} = {v_0^2 \over \Lambda } \tilde{Y}_{ij} \; .
\label{psiy}
\end{eqnarray}
Due to the $Z_2^{}$ discrete flavor symmetry,  the lightest  $\psi_L$ is a massive stable particle and, thus, can be the cold  dark matter candidate. 

\begin{figure}[t]
\includegraphics[width=3cm,height=2cm,angle=0]{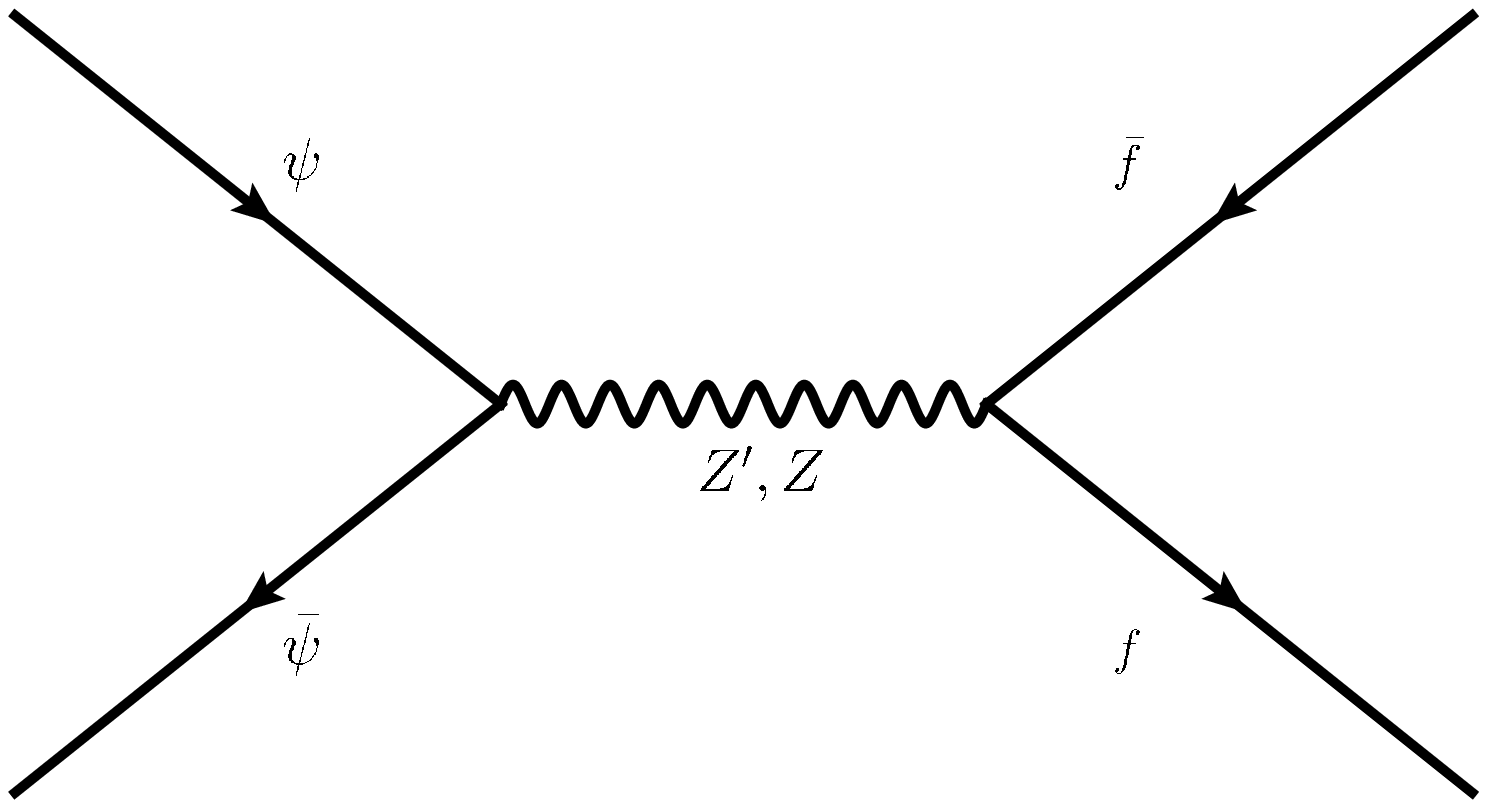}
\hspace{0.5cm}
\includegraphics[width=3cm,height=2cm,angle=0]{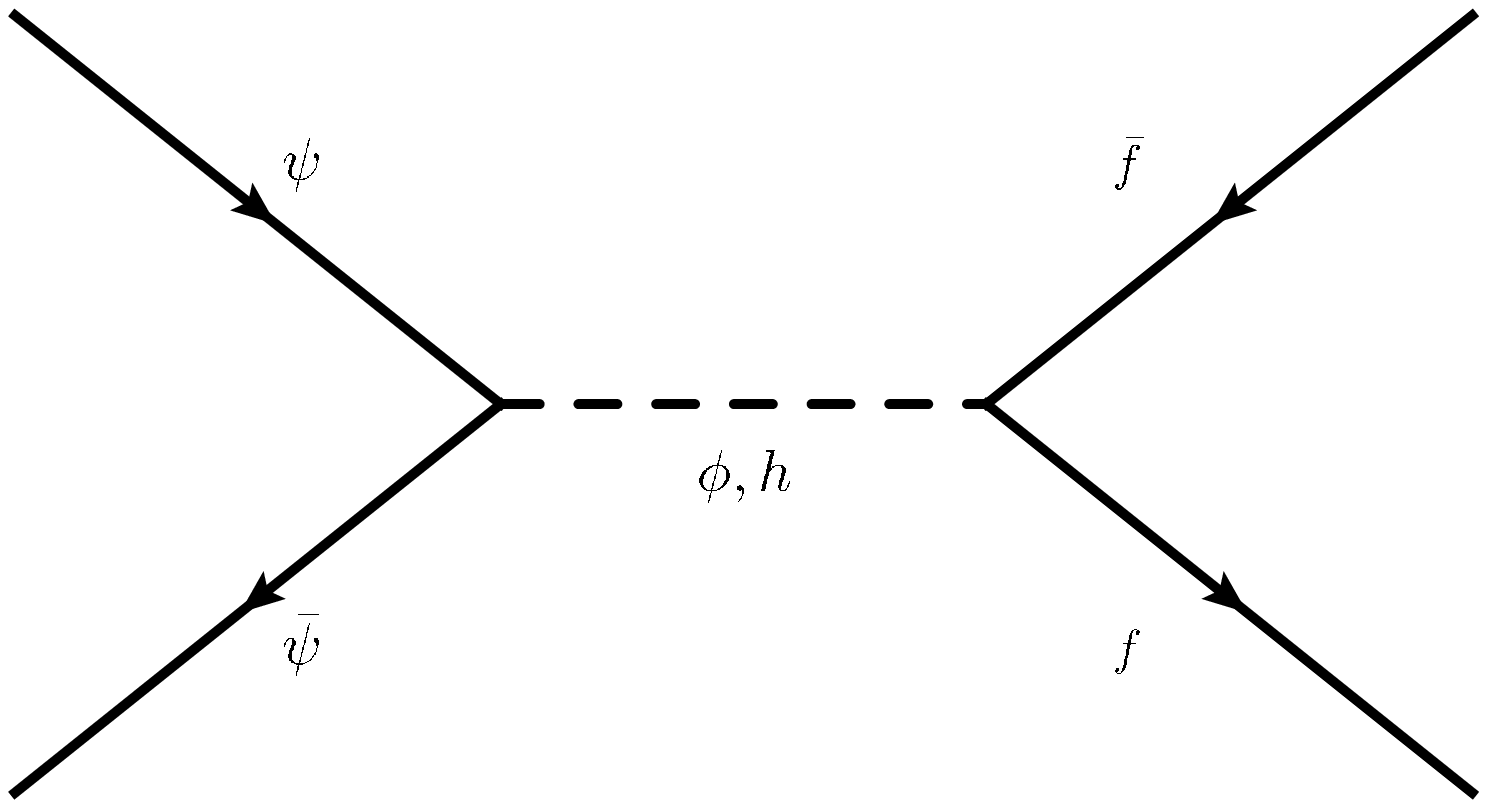}
\hspace{0.5cm}
\includegraphics[width=3cm,height=2cm,angle=0]{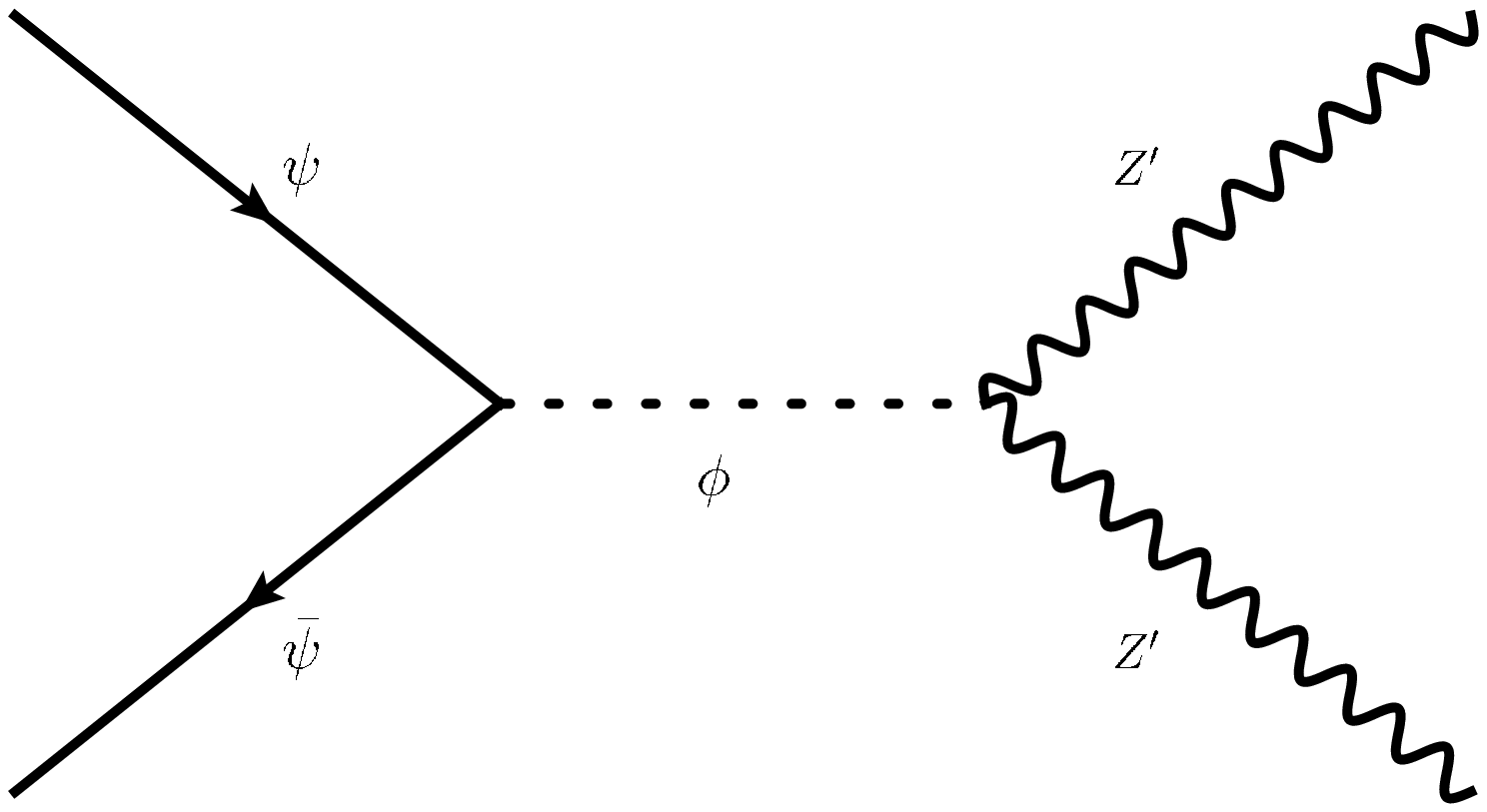}
\hspace{0.5cm}
\includegraphics[width=3cm,height=2cm,angle=0]{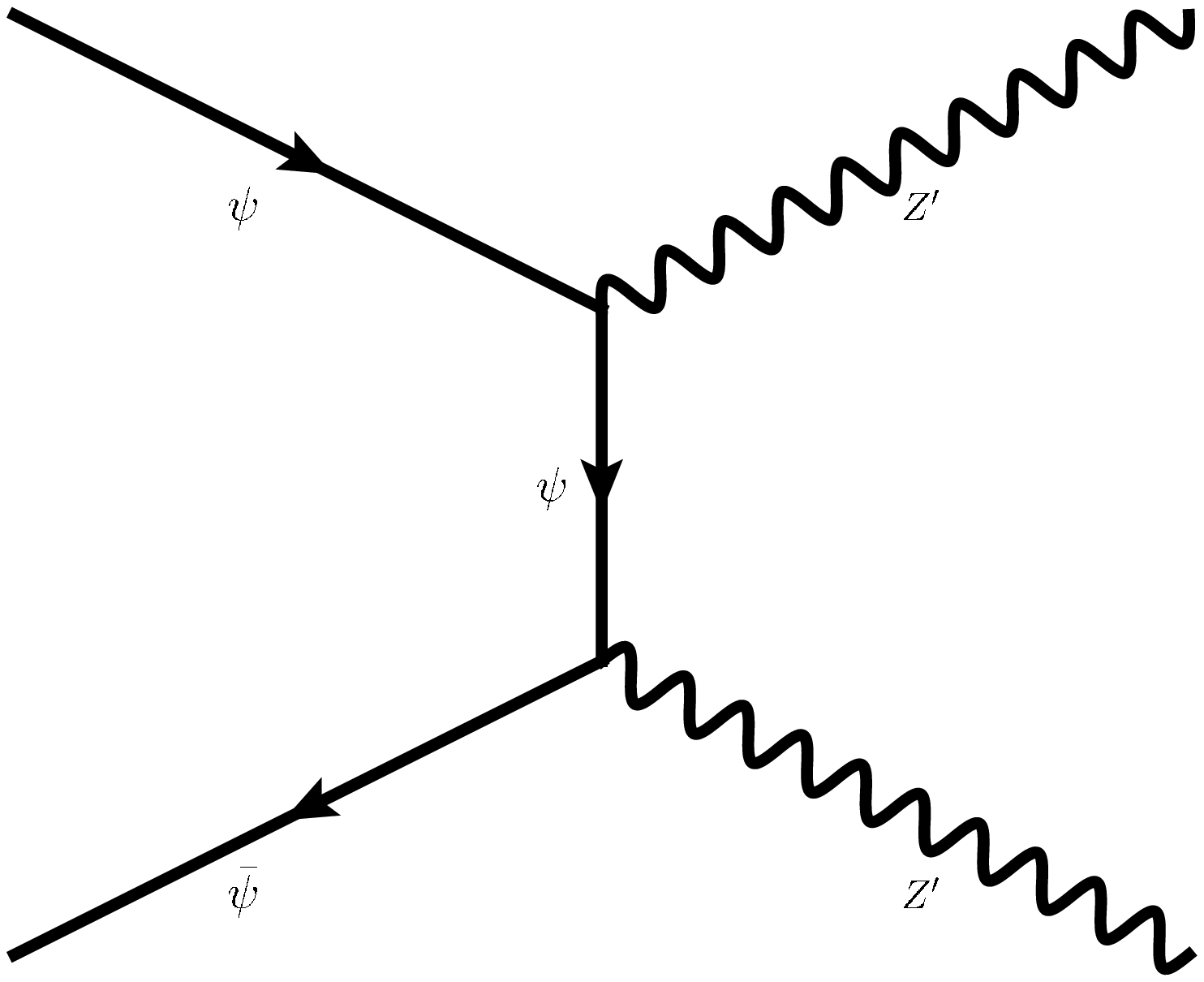}
\caption{Feynman diagrams for the annihilation of the dark matter.} \label{Feyn}
\end{figure}
There are three annihilation channels for the dark matter  illustrated in Fig. \ref{Feyn}: (1) $\psi \bar \psi$ annihilates into  the SM fields through the mixing between $Z$ and $Z^\prime$; (2)  $\psi \bar \psi$ annihilates into the SM particles directly through the mixing between the SM Higgs and the scalar singlet; (3) $\psi \bar \psi$  annihilates into the $Z^\prime Z^\prime$, which decays subquestly into the SM fields through the mixing with the $Z$ boson.  We calculate the resulting relic density and the related direct detection cross section using the Micromegas~\cite{Belanger:2010pz,Belanger:2008sj}, which solves the Boltzman equation numerically and utilizes CALCHEP~\cite{Belyaev:2012qa} to calculate the relevant cross sections.

\begin{figure}[t]
\begin{center}
\includegraphics[width=7.5cm,height=6cm,angle=0]{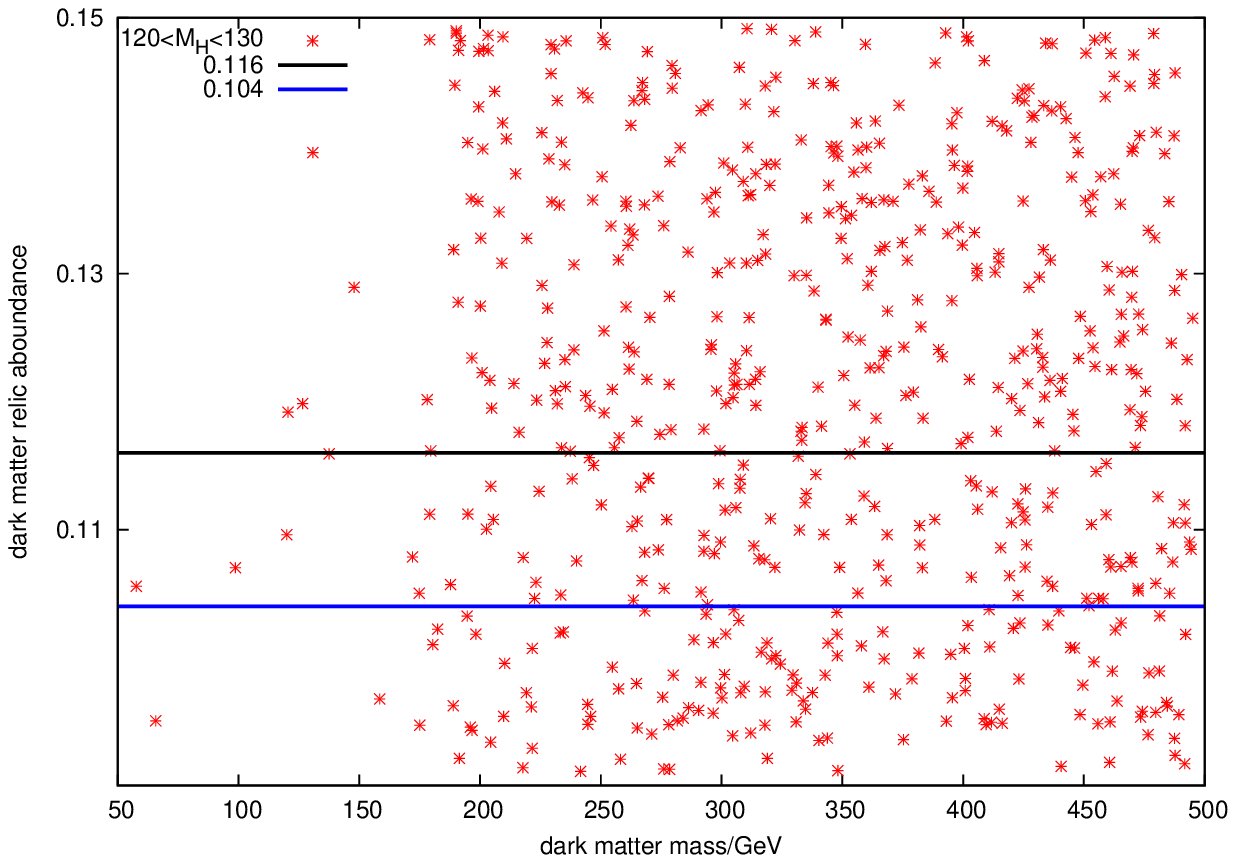}
\includegraphics[width=7.5cm,height=6cm,angle=0]{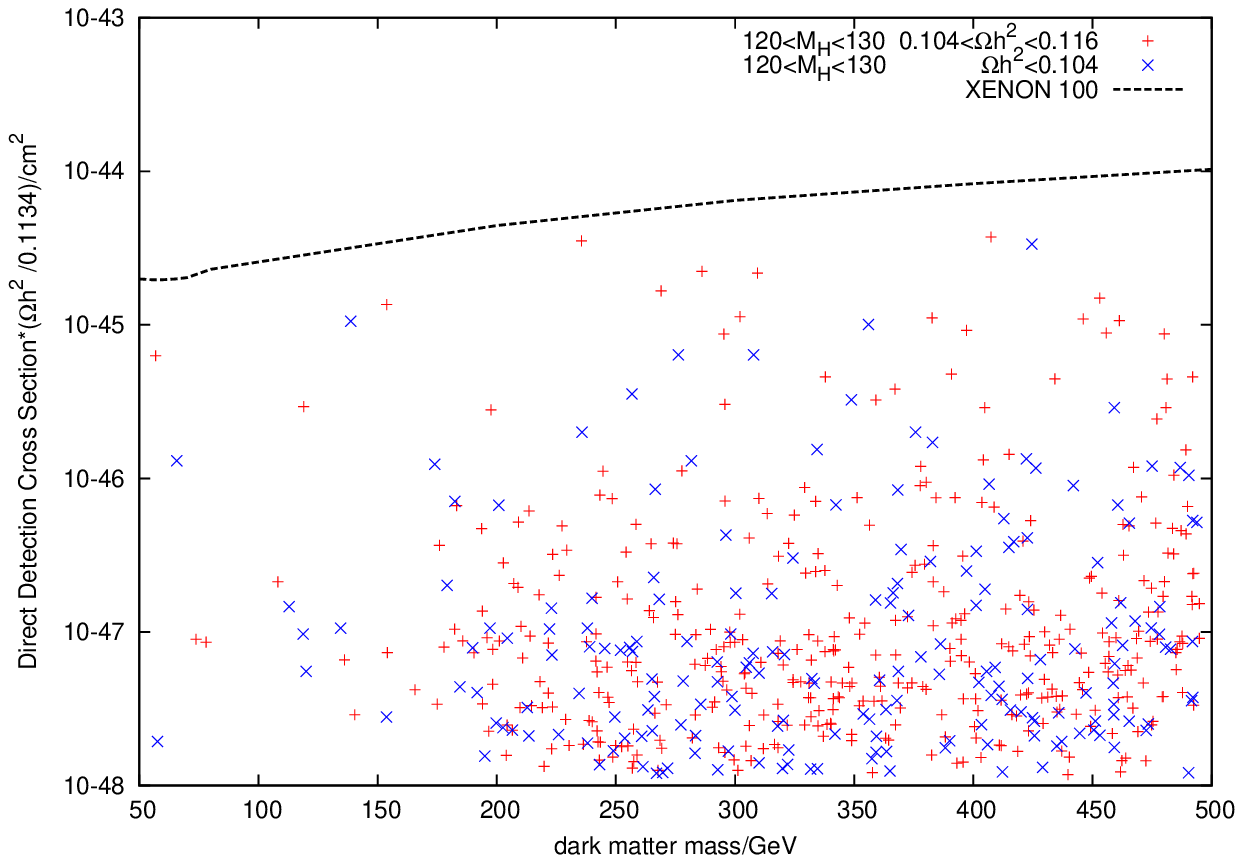}
\end{center}
\caption{Left panel: Dark matter relic abundance as a function of the dark matter mass.  The horizontal band indicates the region consistent with the current relic density measurement from WMAP~\cite{wmap}. Right panel: Scaled dark matter direct detection cross section as a function of the dark matter mass. The dashed line indicates the current bound from XENON 100~\cite{Aprile:2012nq}.} \label{relic}
\end{figure}

We plot in Fig. \ref{relic} (left panel) the dark matter relic abundance as a function of the dark matter mass while  varying the parameters in the scalar potential over the following ranges: $m_0 \in [1\times10^2 ~{\rm GeV}, ~1\times 10^4 ~{\rm GeV}]$, $m_1\in [1\times 10^2 ~ {\rm GeV}, ~ 1\times 10^3~{\rm GeV}]$,  $\lambda_0\in [0.1, ~10]$, $\lambda_1$ and $\lambda_5$ $\in [0.01,~ 1]$, and $M_{Z^\prime} \in [200 ~{\rm GeV}, 500~{\rm GeV}]$. The horizontal band represents the region consistent with the current relic density measurement from WMAP, $0.104< \Omega h^2 < 0.116$~\cite{wmap}. It is evident from the figure that one can achieve the observed relic density  with dark matter mass in the range  $50\sim 500$ GeV.

We plot in Fig. \ref{relic} (right panel) the dark matter direct detection cross section,  scaled by the fraction of the relic density produced, as a function of the dark matter mass. The black dashed line gives the latest experimental constraint from the XENON 100~\cite{Aprile:2012nq}. We conclude that the expected spin-independent direct-detection scattering cross section  is below the current experimental bound. However, one could expect a signal in the future XENON1T\cite{Aprile:2012zx}, which aims to probe the  cross sections  of order $\sigma\sim 2 \times 10^{-47}$ ${\rm cm}^2$, within two years of operation. 

\begin{figure}[h]
\begin{center}
\includegraphics[width=7.5cm,height=6cm,angle=0]{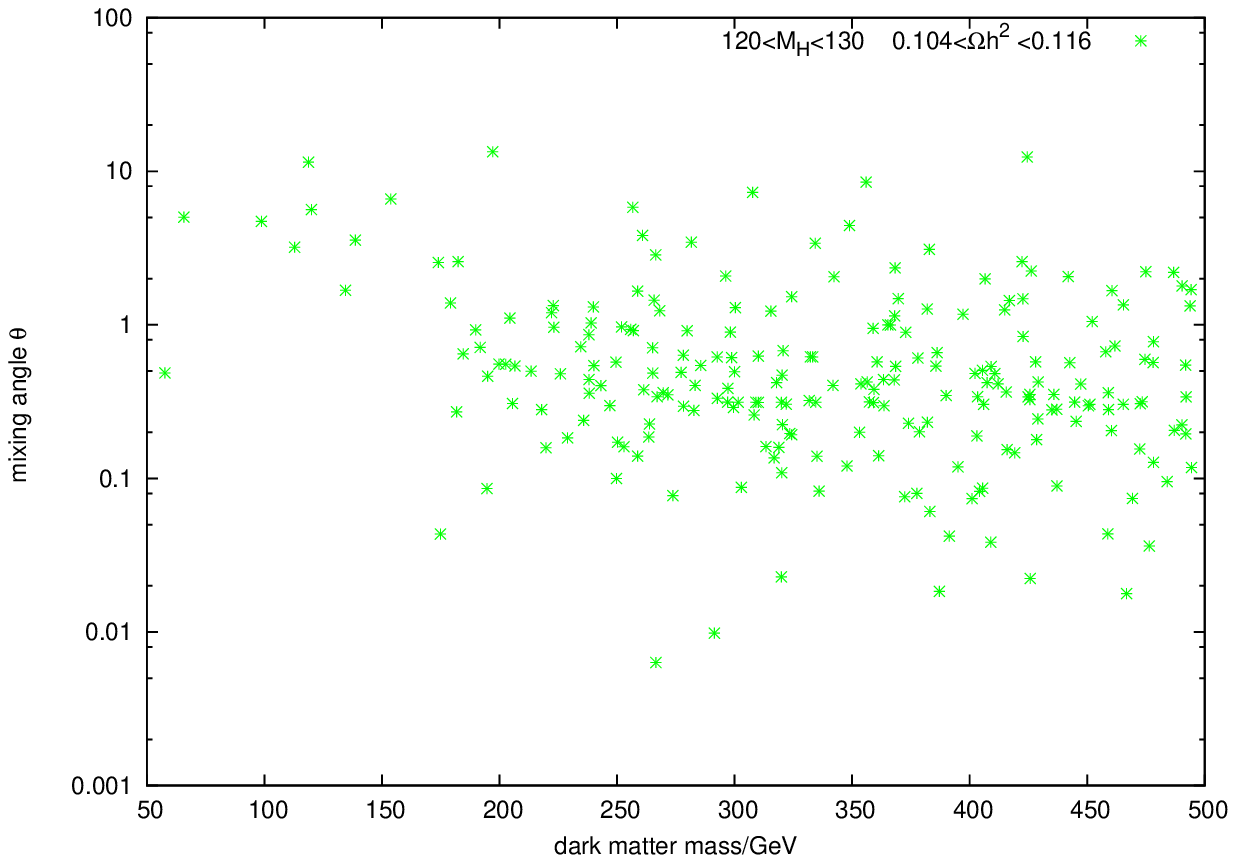}
\includegraphics[width=7.5cm,height=6cm,angle=0]{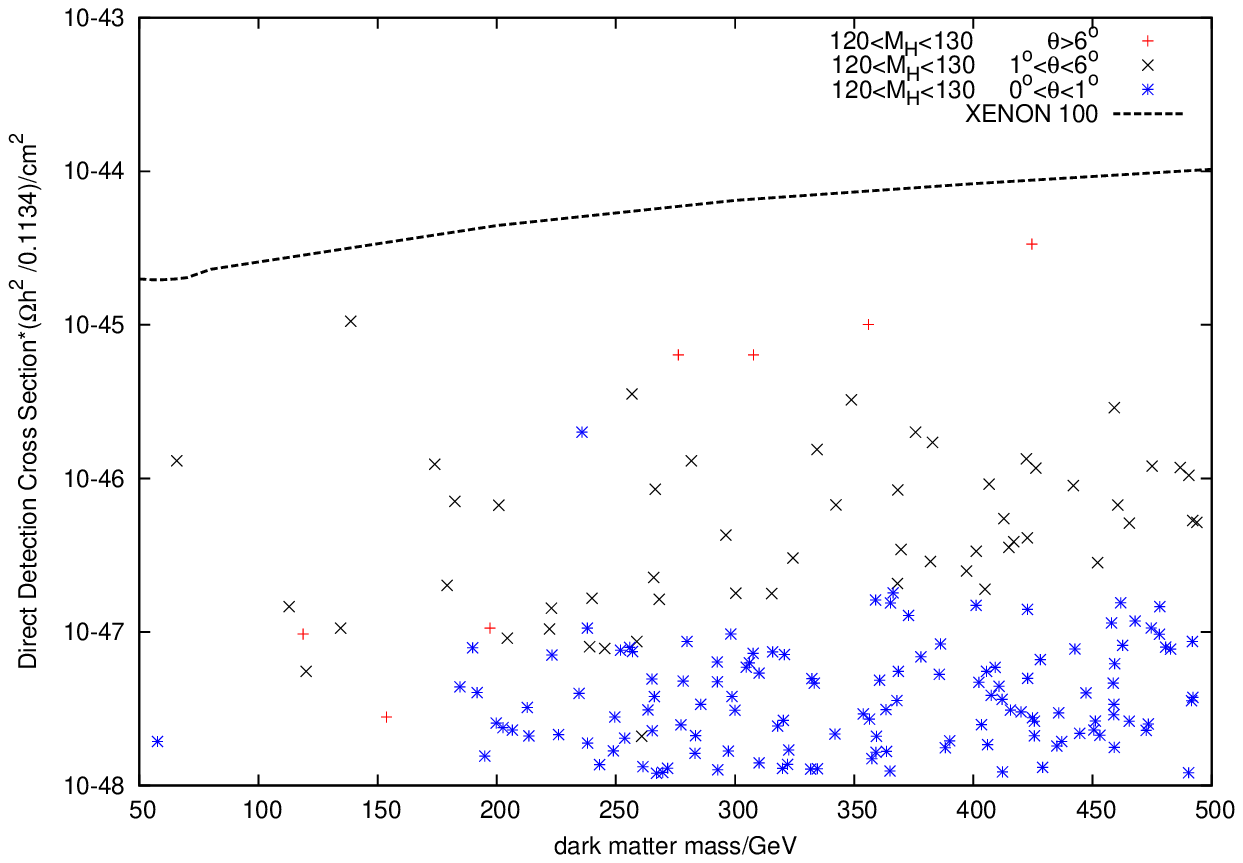}
\end{center}
\caption{Left panel:  $h-\phi$ mixing angle $\theta$ as a function of the dark matter mass.  Right panel: Scaled dark matter direct detection cross section as a function of the dark matter mass, assuming a relic density consistent with WMAP determination. The dashed line the current bound from XENON 100~\cite{Aprile:2012nq}. } \label{direct}
\end{figure}
It is interesting to analyze the dependence of the relic density and direct detection cross section on the $\phi$-$h$ mixing angle $\theta$. To that end, we plot in Fig. \ref{direct} (left panel) the mixing angle as a function of the dark matter mass, assuming saturation of the relic density. We observe that in general the magnitude of the mixing angle is less than 15$^\circ$, making the impact on Higgs boson production at the LHC marginal. In the right panel, we show the scaled direct detection cross section as a function of the mixing angle and dark matter mass. We see that a non-vanishing signal in  future search with a sensitivity of $\sigma_\mathrm{SI-scaled}\sim 10^{-47}$ would require a mixing angle larger than one degree in magnitude. In principle, it is conceivable that a future high-precision Higgs factory  could provide a complementary probe of doublet-singlet mixing at this level. 

\section{Leptogenesis}
In addition to providing a viable dark matter candidate, the $U(1)^\prime$ scenario allows for a viable baryogenesis {\em via} leptogenesis scenario. As the Majorana masses are in the hundreds of GeV range, there exists at least some possibility of testing or constraining this leptogenesis mechanism experimentally. To explore this possibility in an illustrative case, 
we assume that right-handed Majorana neutrinos are hierarchical,
$M_1^{} \ll M_{2,3} $, so that studying the evolution of the number
density of $N_1^{}$ is sufficient.  In the mass scale regime of interest here,
the interactions mediated by all charged lepton Yukawas are  in
equilibrium. Consequently, we should consider flavor-dependent leptogenesis. The
CP-violating asymmetries generated by $N_1^{}$ decays are \cite{lepm1,lepm2}
\begin{eqnarray}
\varepsilon_{\alpha \alpha}^{} = {1 \over 8 \pi } {1
\over(\lambda^\dagger \lambda )_{11}^{} } \sum_j^{} {\rm Im }\left
\{  \lambda^{*}_{\alpha  1} (\lambda^\dagger \lambda)_{1j}
\lambda^{}_{ \alpha j}\right\} g\left( {m_j^2 \over m_1^2} \right) \;
,
\end{eqnarray}
where the wave-function plus vertex contributions are included in \cite{lepm1,lepm2,matterlep}
\begin{eqnarray}
g(x) = \sqrt{x} \left[ {1 \over 1 -x } +1 - (1+x) \ln \left( {1 + x
\over x }\right) \right] \rightarrow -{3\over 2} x^{-1/2} - {5 \over 6} x^{-3/2} ~~~(1 \ll x )\; .
\end{eqnarray}
In addition to $\varepsilon_{\alpha \alpha }^{} $, the final baryon asymmetry depends on wash-out parameters:
\begin{eqnarray}
K_{\alpha \alpha}^{}\equiv{\Gamma (N_1^{} \rightarrow \eta
\ell_\alpha) \over H(M_1^{})} = \frac{|\lambda_{\alpha 1}|^2 M_1/4\pi}{\sqrt{g^\ast} M_1^2/M_\mathrm{pl}}
\equiv {\tilde m_{\alpha \alpha}^{}
\over \tilde m^*} \; ,
\end{eqnarray}
where $H(M_1^{})$ denotes the value of the Hubble rate evaluated at
a temperature $T=M_1^{}$, $M_\mathrm{pl}$ is the Planck mass, $M_1^{}$ is  the mass of the lightest right-handed neutrino, $\tilde m^* =3\times 10^{-3} $ ${\rm eV}$
and $\tilde m_{\alpha \alpha}^{} = { \lambda^*_{\alpha 1} \lambda_{ \alpha  1}^{}
v^2_1  M_1^{-1} }$. Note that we have expressed $K_{\alpha\alpha}$ in terms of a scale associated with the neutrino mass, ${\tilde m}_{\alpha\alpha}$,  and the remaining dimensional factors into ${\tilde m}^\ast$.

The washout factor $K_{\alpha \alpha}^{}$ should be smaller than some maximum,  $K_{\rm Max}$; otherwise the washout effect would be too strong to generate the proper matter-antimatter asymmetry.   For  conventional thermal leptogenesis 
 $K_{\rm Max}\sim1$  and  while it may be as large is $\sim 1000$  for resonant leptogenesis\cite{pilaftsis1}.  As we show below,  $K_{\alpha \alpha}$ can be of ${\cal O}(100)$ the present instance, which is 
since $\varepsilon_{\alpha\alpha}^{}$ in this model can be much larger than that of  the conventional thermal leptogenesis case. 
The lepton asymmetry for the flavor $\alpha$ can be
expressed approximately as\cite{matterlep}
\begin{eqnarray}
\label{eq:yaa}
Y_{\alpha \alpha }^{} \approx {\varepsilon_{\alpha \alpha}^{}\over g_*^{}} \left[\left({\tilde m_{\alpha\alpha} \over 8.25\times 10^{-3}~ {\rm eV}}\right)^{-1} + \left( {0.2\times 10^{-3} ~{\rm eV} \over \tilde m_{\alpha\alpha}}\right)^{-1.16}\right]^{-1} \; . \label{aaaaa}
\end{eqnarray}
Non-perturbative sphaleron interactions partially convert
this lepton asymmetry into a net baryon number asymmetry. Taking
into account the flavor effects, the final baryon asymmetry is given
by\cite{matterlep}
\begin{eqnarray}
Y_B^{} \approx - {8 \over 25 } \left ( {40 \over 13} Y_{ee}^{} + {51
\over 12} Y_{\mu\mu}^{} + {51 \over 13} Y_{\tau\tau}^{}\right)\; .  \label{bbb}
\end{eqnarray}
It should be noted that analytical solutions given in Eqs. (\ref{aaaaa},\ref{bbb}) only work for the case that the $\Delta L=1$ washout effect is much larger than the $\Delta L=2$ wash-out effect, which can be expressed as~\cite{Buchmuller:2004nz}
\begin{eqnarray}
K_{\Delta L=2} \sim  \sum_{ij} |(\lambda^\dagger \lambda)_{ij}^2 |{1 \over M_i M_j} 
\times {T M_{pl} \over 32 \pi^3 \zeta(3) }  \sqrt{ 90 \over 8\pi^2 g^* } \;  ,
\end{eqnarray}
for $T<M_1$. If $K_{\Delta L=2}$ is comparable with the $K_{\alpha \alpha}$, one must resort to a numerical, rather than analytic, solution to the Boltzmann equations.

We emphasize that the neutrino Yukawa couplings enter both $\varepsilon_{\alpha\alpha}$ and the washout factor, translated into ${\tilde m}_{\alpha\alpha}$ in Eq.~(\ref{eq:yaa}). As a result, the magnitudes of the $\lambda_{\alpha 1}$ are strongly bounded: $|\lambda_{\alpha 1}| \lsim 10^{-6}$. If the structure of the neutrino Yukawa matrix is anarchical, we would then expect the magnitudes of the couplings that enter CLFV observables to be $|\lambda_{\mu i}^\ast\lambda_{e i} | \lsim 10^{-12}$, far below a level that could be probed in the next generation of CLFV searches (see Fig.~\ref{fig0}). Consequently, observable CLFV could arise only in the presence of a strongly hierarchical Yukawa matrix, with $|\lambda_{\mu i}^\ast\lambda_{e i} |$ for $i=2,3$ six to seven orders of magnitude larger than for $i=1$. Such a strong deviation from the assumption of anarchy would suggest the presence of an additional lepton flavor symmetry, as we now discuss.

To illustrate how a flavor symmetry might lead to a hierarchical structure, we assume the neutrino sector admits a $U(1)_\ell$ symmetry \cite{pilaftsis}. To be concrete, the $U(1)_\ell$ charges of the fields are given by $Q(\ell_L)=Q(\ell_R) =1$, $Q(1/\sqrt{2}(e^{i \alpha} N_{R2}^{}+ e^{i\beta } N_{R3}^{}) )= -Q(1/\sqrt{2}(e^{-i \alpha} N_{R2}^{}- e^{-i\beta } N_{R3}^{}))=1$ and $Q(N_{R1}^{})=0$. We assume that the $U(1)_\ell$ symmetry is explicitly broken by the GUT or Planck scale physics, such that Yukawa interaction $\varepsilon \overline{\ell_L} H_n^{} N_{R1}^{}$ emerges. In this case, the Yukawa interaction matrix can be given by 
\begin{eqnarray}
\label{eq:yukawamatrix}
\lambda  =\left(  \matrix{\varepsilon & a e^{ i \alpha}&  a e^{i \beta} \cr \varepsilon & b e^{i \alpha} & b e^{i \beta}  \cr  \varepsilon &c e^{i \alpha} & c e^{\beta} }\right) \; ,
\end{eqnarray}
where $\varepsilon$ arises from lepton-number-violating Yukawa interaction term and is, thus, relatively small; $a$, $b$ and $c$ are arbitrary complex parameters, the scale of which is restricted by lepton-flavor-violating decays.
Following from  Eq. (\ref{eq:yukawamatrix}),  the mass matrix of the right-handed neutrinos is given by 
\begin{eqnarray}
M_R^{} = \left( \matrix{M_1^{} & 0&0 \cr 0 & M_2 & M_2 c_{\alpha -\beta} \cr 0 & M_2 c_{\alpha-\beta } & M_2} \right)  + \Delta M\; , \label{SMR}
\end{eqnarray}
where $c_{\alpha-\beta} =\cos(\alpha-\beta)$. Given the neutrino Yukawa coupling  in Eq. (\ref{eq:yukawamatrix}) and heavy neutrino mass matrix in Eq. (\ref{SMR}),  the structure of the  Pontecorvo-Maki-Nakagawa-Sakata (PMNS) matrix implied by neutrino oscillation data  cannot arise solely from the neutrino sector. We recall that the PMNS matrix follows from the mismatch between the diagonalizations of the neutrino mass matrix and the charged lepton mass matrix, {\em i.e.} $V_{\rm PMNS } = V_e^\dagger V_\nu$, where $V_e$ and $V_\nu$, respectively, rotate the left-handed charged and neutral lepton flavor eigenstates to the mass eigenstates. The correct
correct PMNS matrix may then emerge from an appropriately-chosen structure for the charged lepton Yukawa matrix.
\begin{figure}[h]
\begin{center}\includegraphics[width=11cm,height=8
cm,angle=0]{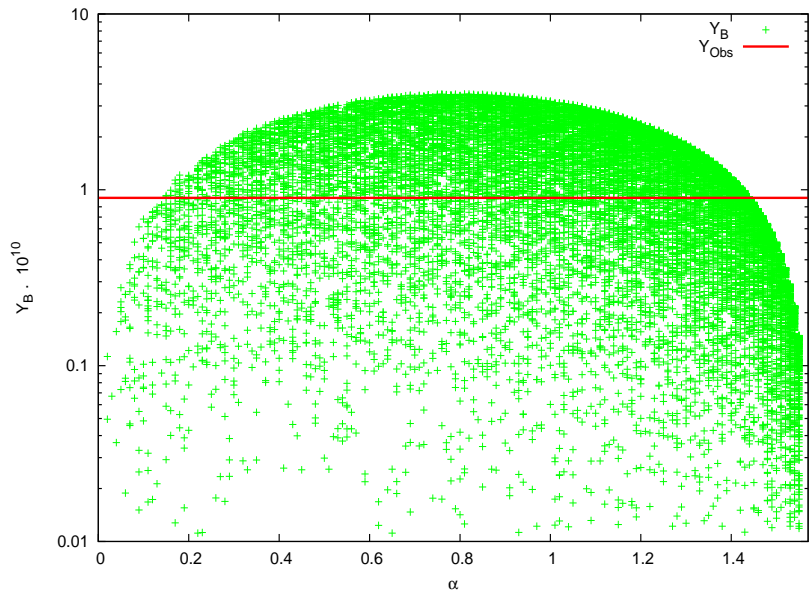}
\end{center}
\caption{$Y_B$ as a function of $\alpha$. The horizontal line  is the center value of the experimental observation.} \label{fig2}
\end{figure}

In Fig.~\ref{fig2} we plot $Y_B^{}$ as a function of $\alpha$, by setting $\varepsilon \sim 10^{-6}$,  $\beta=0$ and $|a|=|b|=|c|\cdot 10^{-2}=10^{-4}$ with their phases vary arbitrarily in the range $[0, \pi]$.  We also set $M_1^{} =200 ~{\rm GeV}$ and $M_2^{} = 500 ~{\rm GeV}$.    In this case, the $\Delta L =2$ washout factor is $ K_{\Delta L=2}^{\rm Max} \sim  0.1  $, which can be neglected compared with the $\Delta L =1 $ wash-out factor: $K_{\alpha\alpha } \sim 50 $.  In this case, we may safely take Eq.~(\ref{aaaaa}) as the analytical solution to the Boltzmann equations. The horizontal line in the figure represents the current experimental value of $Y_B^{}$.
It is clear that the matter-antimatter asymmetry of the Universe can be generated.  Notice that $K_{\alpha \alpha}^{} \gg1$  in our case, which a salient feature of this scenario compared with the conventional thermal leptogenesis case but not one requiring near degeneracies in the heavy neutrino spectrum. Note that for the choice of parameters given here, $|\lambda_{\mu i}^\ast\lambda_{e i}| < 10^{-8}$ for $i=2,3$, implying an unobservable signal in the next generation of CLFV searches\footnote{ More generally, we have estimated that for $ |\lambda_{\alpha i}^\ast\lambda_{\beta i}| \lsim  10^{-6}$ ($\alpha,\beta=e,
\mu,\tau$ and $i=2,3$)  $\Delta L=1$ washout  processes dominate over $\Delta L=2$ processes.  }.

\section{Conclusion}

The search for new physics at the TeV scale is motivated in part by naturalness considerations and in part by the possibility that new TeV scale dynamics may account for the origin of the visible and dark matter of the universe. On the other hand, the observation of neutrino oscillations and the tiny scale of neutrino masses point to new physics at much higher scales, as suggested by the conventional seesaw paradigm. It is interesting to ask whether nature may have generated both neutrino mass and the matter content of the universe at the TeV scale. If so, then one would anticipate signatures in experiments sensitive to BSM physics at this scale. 

In this paper, we have analyzed a simple BSM scenario that can account for dark matter, baryogenesis, and neutrino mass with new TeV scale degrees of freedom and shown that, nonetheless, the experimental signatures are likely to be sparse at best. We have made no attempt to alleviate the Higgs mass fine tuning problem, though it is possible that an embedding of this scenario in a U.V. complete model may do so. In this minimal scenario, a neutrinophilic 2HDM with a local U(1)$^\prime$ symmetry, the second Higgs doublet is entirely responsible for neutrino mass. Its VEV can be naturally small, allowing for $\mathcal{O}(1)$ Yukawa couplings, TeV-scale right-handed neutrinos, and the possibility of observable CLFV. The associated flavor-dependent low-scale (non-resonant) leptogenesis can account for the cosmic baryon asymmetry, while the fermions needed for anomaly cancellation provide a suitable dark matter candidate. 

Despite having the ingredients for a plethora of experimental signatures, we find that simultaneously solving the neutrino mass, dark matter, and baryon asymmetry problems implies that most distinctive features of this scenario (apart from a non-distinctive spin-independent direct detection signal) would be out of reach in the foreseeable future in the absence of additional lepton flavor symmetries. This situation contrasts with a variety of other BSM scenarios, for which new symmetries must be imposed in order to suppress otherwise large deviations from the SM that are inconsistent with observation. In short, nature's solutions to some of the key problems at the interface of particle physics and cosmology may lie at the TeV scale, yet either remain hidden from view or point to an even more complex flavor problem.

\begin{acknowledgments}
We thank O. Seto for alerting us to the potential importance of $\Delta L=2$ washout processes.
This work was supported in part by DOE contract DE-FG02-08ER41531  and the Wisconsin Alumni Research Foundation. 

\end{acknowledgments}

\appendix

\section{$Z-Z^\prime$ mixing at the one-loop level}

The one-loop contributions to the  $\Pi_{ZZ^\prime}^T$ are given by
\begin{eqnarray}
{\Pi}_{ZZ^\prime}^{T 1} &=&+ {  g g^{\prime \prime } (c^2_w -s^2_w) \over (4\pi)^2 c_w^{}} \left\{ \left(-{1\over 6} q^2 + m_\rho^2 \right) (\alpha_\varepsilon^{}+1 ) + q^2 F(m_\rho^2, m_\rho^2,  q^2 )  -2 m_\rho^2 F_1^{} (m_\rho^2 , m_\rho^2, q^2 )\right\}  \nonumber \\ &&+  { g g^{\prime \prime }\over (4\pi)^2 c_w^{}} \left\{  \left(  -{1\over 6} q^2 + {1\over 2} (m_\eta^2 + m_\delta^2 )\right)(\alpha_\varepsilon^{} +1 )\right.  \nonumber \\  && + \left. q^2 F ( m_\eta^{2}, m_\delta^2, q^2 ){\over} -m_\eta^2 F_1^{} (m_\delta^2, m_\eta^2, q^2) -m_\delta^2 F_1^{} (m_\eta^2 , m_\delta^2 , q^2 )\right\} \; , \\
\Pi_{ZZ^\prime}^{T2} &=&- {c_w^{} g g^{\prime 3}  v_2^2  \over 2 (4\pi)^2} \left[{\over}\alpha_\varepsilon^{} - F_0^{} (m_W^{2}, m_\rho^2, q^2 )\right]+ {g^3 g^{\prime \prime } v_2^2 \over 2 c_w^3 (4\pi)^2}\left[{\over } \alpha_\varepsilon^{} - F_0^{} (m_Z^2, m_\eta^2 , q^2) \right] \nonumber \\ && + { g^{} g^{\prime \prime 3} v_2^2 \over 2 c_w^{} (4\pi)^2} \left[ {\over }  \alpha_\varepsilon^{} - F_0^{}(m_{Z^\prime}^2, m_\eta^2, q^2 )\right] \; , \\
\Pi_{ZZ^\prime}^{T3} &=&- { g g^{\prime \prime }  \over 2 c_w^{} (4 \pi)^2}\left\{ (m_\eta^2 + m_\delta^2) (\alpha_\varepsilon^{} +1) {\over } -m_\eta^2 \ln m_\eta^2 -m_\delta^2 \ln m_\delta^2 \right\}\nonumber   \\ &&-{ g g^{\prime \prime }  (c_w^2 -s_w^2 ) m_\rho^2 \over c_w^{} (4\pi)^2} \left({\over } \alpha_\varepsilon^{} + 1 -\ln m_\rho^2  \right) \; ,
\end{eqnarray}
where $\alpha_\varepsilon = 1/\varepsilon -\gamma_E + \ln 4 \pi + \ln \mu^2 $ and $g^{\prime \prime}$ is the coupling constant of the new U(1) gauge symmetry.  $\Pi_{ZZ^\prime}^{T1}$ comes from the vertex with two Higgs bosons and one gauge field,  $\Pi_{ZZ^\prime}^{T2}$ comes from the vertex with two gauge fields and one Higgs boson, while  $\Pi_{ZZ^\prime}^{T3}$ comes from the  vertex with two gauge fields and two Higgs bosons.   We define $\hat \Pi_{ZZ^\prime}^T\equiv\Pi_{ZZ^\prime}^{T1}+\Pi_{ZZ^\prime}^{T2}+\Pi_{ZZ^\prime}^{T3}$.

\end{document}